\documentclass[aps,pra,showpacs,superscriptaddress,amssymb,floatfix,tightenlines,twocolumn]{revtex4}
\newcommand{\beq}{\begin{equation}}
\newcommand{\eneq}{\end{equation}}
\newcommand{\beqy}{\begin{eqnarray}}
\newcommand{\eneqy}{\end{eqnarray}}
\usepackage{graphics}
\usepackage{times,mathptm}
\usepackage{amsmath}
\usepackage{epsfig}
\usepackage{dcolumn}
\usepackage{bm}
\usepackage{mathrsfs}
\usepackage{theorem}
\usepackage{amsmath,amssymb}
\usepackage[colorlinks]{hyperref}

\newcommand{\qed}{\nobreak \ifvmode \relax \else
\ifdim\lastskip<1.5em \hskip-\lastskip \hskip1.5em plus0em
minus0.5em \fi \nobreak \vrule height0.75em width0.5em
depth0.25em\fi}

\newcommand{\ket}[1]{\left| #1 \right\rangle}
\newcommand{\bra}[1]{\left\langle #1 \right|}

\begin{document}

\title{Globally controlled universal quantum computation with arbitrary subsystem dimension}
\author {Gerardo~A. Paz-Silva}
\altaffiliation{E-mail: {\tt gpazsil@ics.mq.edu.au}}
\author {Gavin K. Brennen}
\author {Jason Twamley}
\affiliation{Centre for Quantum Computer Technology, Macquarie University, Sydney, NSW 2109, Australia}

\date{\today}

\begin{abstract}
We introduce a scheme to perform universal quantum computation in quantum cellular automata (QCA) fashion in arbitrary subsystem dimension (not necessarily finite). The scheme is developed over a one spatial dimension $N$-element array, requiring only mirror symmetric logical encoding and global pulses. A mechanism using ancillary degrees of freedom for subsystem specific measurement is also presented.  \end{abstract}


\maketitle
Quantum computation involves control at some level of a large number of quantum subsystems. The traditional circuit based approach is to have a set of subsystems for storing quantum information, denoted q-sites, which are fully addressable not only individually but in subsets~\cite{general}. It is known that universal quantum computation is possible, independent of subsystem dimension, when arbitrary single q-site unitaries and 2-q-site entangling gates are available~\cite{Brylinski, LloydBraunstein}. Yet addressing individual subsystems in large arrays is extremely challenging and can impose significant errors due to miss-alignment of the fields which unintentionally act on neighbors to the target (cross-talk) or miss the target.  Thus as a mean to avoid it another option appeared: global control schemes~\cite{globalQC, Raussendorf1, FitzsimonsTwamley}. 

The philosophy behind global control is to reduce the interaction with the array of q-sites to require only global manipulation, implemented for instance through global fields homogeneously coupled with all q-sites. In general such global control schemes require a natural evolution (time step) for the array, and tailored sequences of global pulses which translate physical asymmetries in the array into control of particular sites or, equivalently, chronological control into spatial control.  When the resultant evolution is a set of gates which act on small neighborhoods in parallel, this is also a quantum cellular automata (QCA) model. The models for global control have been so far concerned only with qubits~\cite{Raussendorf1, globalQC}, however with the development of higher dimensional computational models (using qudit computation and continuous variables(CV), also called qunat computation) that show advantages in terms of efficiency and robustness~\cite{higherQC}, the natural direction is to find a way to implement such models in a globally controlled fashion. 

The aim of this paper is to develop such a model in one spatial dimension, inspired by a previous protocol restricted to qubits~\cite{Raussendorf1} and recent results on globally controlled transport of qudits and qunats~\cite{PRTD}. The main technical difficulty is the presence of more complex phases that appear in higher dimensions, in contrast with the $\{1,-1\}$ phase in the qubit case, and solving equations for discrete variables which are defined modulo the dimension $d$ of the logical q-site under consideration. Fortunately, in~\cite{PRTD} we developed most of the tools we will need as well as revising some of the known results available in the literature~\cite{Wang}. 

The paper is organized as follows: in Sec.\ref{1QC} we review some mathematical results on bases for the operator space particularly exploring a Hermitian basis so that we can obtain universal QC through gates generated by physically accessible Hamiltonians.  In Sec.\ref{UQC} we develop our scheme and show how certain sequences of global pulses plus a natural time step, acting on a N-site array with a mirror-symmetric logical encoding, can generate single site arbitrary unitary gates as well as nearest neighbor entangling gates and thus achieve universal quantum computation in a quantum cellular automata fashion.  Input and output of information into the array is discussed in Sec.\ref{IO} and we conclude with a summary of the results.
    
\section{Single qusite QC}
\label{1QC}

The state spaces for discrete and continuous q-sites are qualitatively different as unitary operators on these spaces are generated by the algebras $\mathfrak{su}(d)$ and polynomials in $\{\hat{q},\hat{p},{\bf 1}\}$ respectively where the canonical commutation relation $[q,p]=i\hbar$ is satisfied.  Motivated by results in Ref. \cite{Vlasov2} we can chose a set of generators for single q-site quantum gates using Hermitian counterparts of Weyl pairs which satisfy the the commutation relations of the generalized Pauli group for qudits and the Weyl representation of the Heisenberg commutation relation for CVs.

\subsection{The qudit case ($d$ finite)}

The state space for a single qudit is $\mathcal{H}_d={\rm span}_{\mathbb{C}}\{\ket{j}\}_{j=0}^{d-1}$ and since the global phase is irrelevant, unitary gates are elements of the group $SU(d)$.
A unitary operator basis on this space is $\{ X(a)Z(b); a,b\in \mathbb{Z}_d \}$ where $X=\sum_{s=0}^{d-1}\ket{s+1}\bra{s}$, and $Z = \sum_{s=0}^{d-1}e^{i\frac{2 \pi}{d} s} \ket{s}\bra{s}$ with additional modulo $d$ satisfying $X^d = 1 = Z^d$.  We adopt the notation $X(a)\equiv X^a$ and $Z(b)\equiv Z^b$.  The group commutator is $Z(b) X(a) = e^{\frac{i 2\pi}{d} ab} X(a) Z(b)$.   However if we are to perform quantum gates generated by Hamiltonians we require a Hermitian basis of $\mathfrak{su}(d)$ such as the following,
\begin{equation}
\begin{array}{lll}
B(a,b) &=&   e^{i\phi_{a,b}} X(a)Z(b) + e^{-i\phi_{a,b}} Z(-b)X(-a) \\ 
&=&\cos(\phi_{a,b}) (X(a) Z(b) + Z(-b) X(-a)) \\
&&+ \sin(\phi_{a,b}) i (X(a)Z(b) - Z(-b) X(-a) )
\end{array}
\label{opbasis}
\end{equation}
where $\phi_{a,b}=\pi/4 +  \pi a b / d$. Since the two Hermitian summands commute we can write an arbitrary unitary U $\in SU(d)$ as
\beq
\begin{array}{lll}
U&=& \prod_{a,b=0}^{d-1} e^{i\alpha_{a,b} \cos(\phi_{a,b}) (X(a) Z(b) + Z(-b) X(-a)) }\\
&& e^{\alpha_{a,b} \sin(\phi_{a,b}) (X(a) Z(b) - Z(-b) X(-a)) }
\end{array}
\eneq
where the $\alpha_{a,b}$ are real numbers (note the presence of $d^2-1$ independent real parameters). It follows that if we can perform $ e^{ i \kappa i^{\frac{1\mp1}{2}}  (X(a) Z(b) \pm Z(-b) X(-a))}$, for some $\kappa$, then we are able to construct an arbitrary unitary.

\subsection{The CV case ($ d \rightarrow \infty$)}

The state space for a single qunat is $\mathcal{H}_{CV}=\mathcal{L}^2(\mathbb{R})$.  Consider the following operators on this space:
\beq
\label{ZXoriginal}
Z(\alpha) = e^{i \alpha \hat q}  \,\,\,\,\,{\rm and} \,\,\,\,\, X(\beta) = e^{-i \beta \hat p},  
\eneq
satisfying $Z(\alpha) X(\beta) = e^{i \alpha \beta} X(\beta) Z(\alpha)$ which is the Weyl representation of the Heisenberg group  (having set $\hbar =1$). We can fix a computational basis $\ket{q}$ satisfying the relations $Z(\alpha)\ket{q}=e^{i\alpha q}\ket{q}$ and $X(\beta)\ket{q}=\ket{q+\beta}$.
 Unitaries generated by the set of Hermitian operators corresponding to the real and imaginary parts of $X(\beta)Z(\alpha)$ can be constructed,
 and in Ref. \cite{Vlasov2} it is argued that such gates are universal for single CV computation.  To elucidate this point, consider the unitary evolutions obtained using the gate library
\beq \label{set} \{ e^{i a \cos{{\omega_x} \hat p}}, e^{i b \sin{{\omega_x} \hat p}},  e^{i c \cos{{\omega_z} \hat q}}, e^{i d \sin{{\omega_z} \hat q}} \},\,\,\,\, a,b,c,d \in \mathbb{R}
\eneq we shall argue that this components are enough to have single qunat universal computation i.e. are enough to generate any Hamiltonian evolution. The core of the argument is a result from functional analysis~\cite{FuncAnalysis} which shows that the set $\{\sin{n x}, \cos {n x}\}$ is a basis for the functions space, in particular to the polynomials space.
\beq 
f(x) = C_0 + \sum_{n=1}^{\infty} C_n \cos{n x} + D_n \sin{n x}, \,\,\,\, \forall f(x)
\eneq
It follows then that we can approximate any Hamiltonian in $q$ and $p$ through a convenient sequence of basis elements actions, or products of sums, with arbitrary accuracy. However this ability comes at a cost, we are able to reproduce the function only within a domain $\{-L,L\}$, and then it starts repeating itself effectively making the phase space periodic $\hat q = \hat q + nL$, $\hat p = \hat p + mL'$ for some $L,L'$ and all $n,m$, allowing us to define periodic coordinates $q_L, p_L$. 
\beqy 
\nonumber \hat{q}, \hat{p} &\longrightarrow& \hat{q}_L , \hat{p}_L \\
\nonumber \left[\hat{q}, \hat{p}\right]= i   &\longrightarrow& [\hat{q}_L, \hat{p}_L] = i 
\eneqy
In this way we would have, for instance
\beqy 
\nonumber e^{i \alpha \hat q_L} &=& e^{i \alpha \sum_{n=1}^{\infty} (\frac{- 2 (-1)^n}{n}) \sin{n \hat q_L} } \\
\nonumber e^{i \alpha \hat q^3_L} &=& e^{i \alpha \sum_{n=1}^{\infty} (\frac{- 2 (n^2 \pi^2-6) (-1)^n}{n^3}) \sin{n \hat q_L} }
\eneqy
within $\{-\pi,\pi\}$. 

Since the terms inside the argument of the exponential commute we can apply then a product of exponentials and truncate at some $n_{\rm max}$ to approximate a polynomial in $q$:
\[
e^{i \alpha \hat f(\hat{q}_L)}\approx \prod_{n=1}^{n_{o\rm max}}e^{i\alpha \tilde{f}_o(n)\sin{n \hat q_L}}\prod_{n=1}^{n_{e\rm max}}e^{i\alpha \tilde{f}_e(n)\cos{n \hat q_L}}
\]
where $\tilde{f}_{e(o)}(n)$ is the discrete cosine(sine) Fourier transform of $f(\hat{q}_L)$ \footnote{For example, we find that for $n_{\rm max}=1000$ the error in the approximation $\epsilon=|e^{i \alpha \hat q^3_L}-U_{\rm approx}|/\sqrt{2}$ for $|\alpha q|<\pi$ is $\epsilon<0.018$}. Given the ability to generate an arbitrary  qubic polynomial in $\hat{q}_L$ and a quadric polynomial in $\hat{p}_L$, it is possible using the Lie-Trotter theorem to generate any polynomial $f(\hat{q}_L,\hat{p}_L)$~\cite{LloydBraunstein}. This allows us to have then the same results in the standard phase space, namely Hamiltonians and their respective eigenstates written in terms of the new $q_L,p_L$ coordinates. Alternatively one could directly span the any order polynomial as an expansion of sine and cosines. Thus, in principle, universal single qunat gates with arbitrary accuracy are viable in the $\{-L,L\}$ domain with currently physically achievable Hamiltonians. 
Thus to approximate a general Hamiltonian we must be able to construct Hamiltonians of the form $e^{i(X(a)Z(b) + Z(-b)X(-a))}$ or $e^{i(X(a)Z(b) - Z(-b) X(-a))}$. To do so we start from the elements in ~\eqref{set}, now if we can also perform global $e^{i(q^2 + p^2)\omega}$ rotations we can induce the transformation  \beqy \nonumber\hat q &\rightarrow&\hat{q} \cos\omega  +\hat{p} \sin\omega \\ 
\nonumber \hat{p} &\rightarrow& -\hat{q} \sin\omega  +\hat{p} \cos\omega 
\eneqy
and thus $e^{-i(q^2 + p^2)\omega}e^{i\alpha\cos{(\beta \hat q)}}e^{i(q^2 + p^2)\omega}=e^{i\alpha\cos{(a q + b p)}}$ with $a,b \in \mathbb{Z}$, choosing $\cos\omega = a/\sqrt{a^2+b^2}$, $\sin\omega = b/\sqrt{a^2+b^2}$, and $\beta = \sqrt{a^2 + b^2}$. Similarly, conjugating evolution generated by $\sin(\beta\hat{q})$ we can obtain $e^{i\alpha\sin{(a q + b p)}}$.  Hence using the fact that those evolutions commute 
and that 
$X(a)Z(b) + Z(-b) X(-a) =2 (\Re(e^{i ab}) \cos{(ap +bq)} - \Im(e^{iab}) \sin{(ap + bq)})$ and
$X(a)Z(b) - Z(-b) X(-a) =2 (\Im(e^{i ab}) \cos{(ap +bq)} + \Re(e^{iab}) \sin{(ap + bq)})$
we can construct any unitary 
$U=e^{i\alpha (X(a)Z(b)+Z(-b)X(-a))}$ or $U=e^{\alpha (X(a)Z(b)-Z(-b)X(-a))}$. 
For universal qunat computation, the authors in Ref. \cite{LloydBraunstein} showed that it suffices to have arbitrary single qunat gates and at least linear coupling in $q$ between qunats.  Hence we confirm the argument in Ref. \cite{Vlasov2}. 

In summary, to implement a universal scheme for quantum computation, we will need to be able to apply the set of operators~\eqref{opbasis} on any site $s$, and furthermore we need an entangling gate between at least nearest neighbors $\{s, s+1\}$. In the following section we shall explore a QCA scheme on a $1-D$ lattice with N-sites, where using only global pulses we can achieve the required addressability for arbitrary dimension ($d$) elements. Most of our calculations are equivalent as we are mainly using the algebraic and transformation properties of the $Z(u)X(v)$ operators, thus a unified notation  $Z(u) X(v) = \zeta^{uv} X(v) Z(u)$ where $\zeta = e^{i\frac{2\pi}{d}}$ for finite $d$ and $\zeta = e^{i}$ for CV, is in order.
  
\section{Universal computation with global control}
\label{UQC}

Inspired in the scheme developed by Raussendorf~\cite{Raussendorf1} we generalize it to arbitrary dimension. Anticipating the results of the the paper we shall fix at this point a mirror-invariant encoding of the initial state, effectively turning Eq.~\eqref{ideal} into a single logical site gate. We will encode $M$ logical q-sites into $N$ physical sites, the mirror symmetric encoding leads us to two somewhat similar encodings depending if $N$ is odd or even. If $N$ is odd we can encode $M=\frac{N-1}{2}$ logical q-sites, while if $N$ is even we can encode $M=N/2$ q-sites i.e. $N$ q-sites encode $\lfloor N/2\rfloor $ logical q-sites. More specifically we do this by mapping $\rho_{1,...,M} \rightarrow \rho_{1,...,M} \otimes \chi_{\rm middle} \otimes \rho_{M,...,1}$, where the $\chi_{\rm middle}$ contains no encoded information and is dropped if $N$ is even.


We start recalling the step operator $T$ for arbitrary dimension introduced in~\cite{PRTD} for a $N$-elements chain,  
\[
T = \prod_{j=1}^{N} F^{-1}_j \prod_{i=1}^{N-1}  CZ_{i,i+1}    
\]
where $F_j$ is the Fourier gate acting on element $j$ (satisfying $F^4 = 1$) and $CZ_{i,i+1}$ is the generalized control phase gate between sites $i$ and $i+1$. For qudits $F_j=\frac{1}{\sqrt{d}}\sum_{r,s=0}^{d-1}\zeta^{rs}\ket{r}_j\bra{s}$ and $CZ_{i,i+1} = \sum \zeta^{-ij} Z^i\otimes Z^j$
while for CV, $F=e^{i\frac{\pi}{4}(q^2+p^2)}$ and $CZ_{i,i+1}=e^{i q_i\otimes q_{i+1}}$.  Henceforth,
all operations are global so we write $F=\prod_{j=1}^{N} F^{-1}_j$ and $CZ= \prod_{i=1}^{N-1}  CZ_{i,i+1}    $. 
The composition $F^{2} T^{N+1}$ is a reflection about the middle on the state of an $N$-element chain \cite{PRTD}. Consider the operator,
\beqy
\nonumber\tilde T_{\vec{\epsilon}}&=& P(\epsilon_{N+1})T P(\epsilon_N)T P(\epsilon_{N-1})...T P(\epsilon_0)\\
\nonumber&=& P(\epsilon_{N+1})(T P(\epsilon_N)T^{-1}) (T^{2}P(\epsilon_{N-1})T^{-2})\\
\nonumber&&\ldots(T^{N+1} P(\epsilon_0)T^{-(N+1)})T^{N+1}\\
\nonumber&=&\prod_{m=0}^{N+1} P(\epsilon_{N+1-m})(m) T^{N+1} = \bar{T}_{\vec{\epsilon}} T^{N+1} ,
\eneqy
where $P(\epsilon_s) = \bigotimes_{i=1}^N X_i(-\epsilon_s)Z_i(\epsilon_s)$, and also the following action 
\beqy
\nonumber V(\alpha,u,v,l) &=& \left(F^2 \tilde T_{\vec{\epsilon}}\right)^{-1}\prod_j e^{ -i\, \beta/2 \,(X(u)Z(v) + (X(u)Z(v))^\dagger)} \\
\nonumber&& \left(F^2 \tilde T_{\vec{\epsilon}}\right) \prod_j e^{ -i\, \beta/2 \, (X(u)Z(v) + (X(u)Z(v))^\dagger)}\\
\nonumber&=& \label{pre} \left( F^2 \bar T_{\vec{\epsilon}} F^2\right)^{-1} \prod_j e^{- i \beta/2 (X(u)Z(v) + (X(u)Z(v))^\dagger)} \\
\nonumber&& \left( F^2 \bar T_{\vec{\epsilon}} F^2\right) \prod_j e^{ i \beta/2 (X(u)Z(v)+ (X(u)Z(v))^\dagger)}\\
&=^!& \label{ideal} \prod_j e^{ i \alpha /2 (X(u)Z(v) + (X(u)Z(v))^\dagger) \Delta_{j,l}},
\eneqy
for some $l$, where we have defined $\Delta(j,l)=\delta_{j,l}+\delta_{j,N+1-l}$, the symmetric delta function.  Here we have used the fact that for mirror symmetric operators (which is the only kind used in our control), $F^2T^{N+1}$ acts as the identity.  The overall result is then a unitary generated by $X(u)Z(v)+ (X(u)Z(v))^\dagger$ on site $l$ and its mirror image only. Here where $u,v \in \{0,...,d-1\}$ for the qudit case and $\in \mathbb{R}$ for the CV case. Before proceeding with the calculation it is worthwhile giving some motivation to such claim: if one does the calculation for some short chain, say 6 qudits (or qunats), and sees how a homogenous $P(\epsilon)$ evolves using the time step introduced in~\cite{PRTD} one realizes that every site (and the mirrored position) ends up with a different element of the Clifford group, and thus a Clifford operator commuting with such time-evolved operator will gain a different phase which will depend on the particular site of the chain it is acting. This was first exploited for qubits in~\cite{Raussendorf1}. Note also that given the overall mirror symmetry of our protocol, at most we only need $\epsilon_N,...,\epsilon_{[(N+1)/2]}$ to be non-vanishing.

\begin{figure}
\begin{center}
\includegraphics[width=\columnwidth]{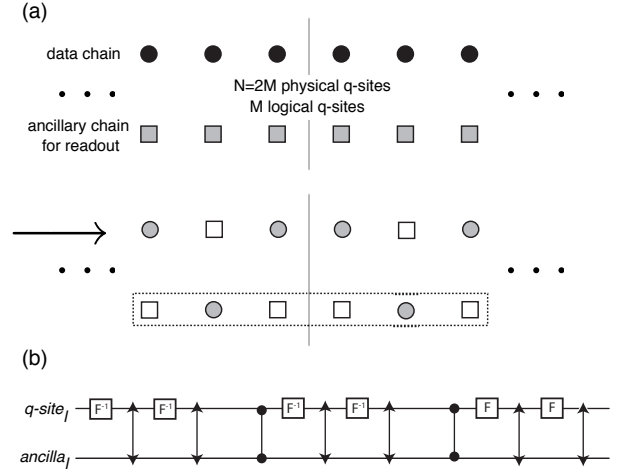}
\label{fig:1}
\caption{(a) The architecture consists of a chain of $N$ physical q-sites encoding $M$ logical q-sites (shown here for $N$ even) in a mirror symmetric pattern, i.e. identifying site $l$ with site $N+1-l$.   To perform readout, an ancillary q-site (another qudit or another CV degree of freedom) is associated with each data q-site shown here as a another parallel chain, though the ancillary and data q-sites may be spatially co-located.  The ancillary q-sites are initialized in $\ket{0}^{\otimes{N}}$. (b) A circuit for executing the SWAP gate between data and ancillary q-sites at locations $l$ and $N +1 -l$. We require the ability to execute a global gate $CZ_{d,a}$ between the two chains, where we recall that the global swap gate can be executed through $CZ$ and global $F$ gates. Also note that the remaining sites, on which no $F$ gate is performed, undergo trivial evolution. Finally, the required sites are swapped and we can perform a global measurement on the ancillary chain to readout logical q-site $l$. The process can be repeated subsequently for the remaining sites. Note that only global operations/measurements are used.}
\end{center}
\end{figure}

To show this explicitly we must perform the direct calculation, so first we introduce a convenient notation for our Clifford operators and their evolution. 

The operator $A(t) = \zeta^{f(t)} \bigotimes_j X^{x_j} Z^{z_j} $ can be written as $A(t) = \zeta^{f(t)} \tau_{\vec{a}}$ with $\vec{a} = (\vec{x}_a, \vec{z}_a) = ( x_{1},...,x_{N}, z_1,...,z_N)$. In this notation, $\tau_{\vec{a}}\,\, \tau_{\vec{b}} = \zeta^{\vec{x}_b \cdot \vec{z}_a- \vec{x}_a \cdot \vec{z}_b} \,\tau_{\vec{b}}\,\, \tau_{\vec{a}}$, and the evolution is given by
\beqy
\nonumber
T: A(t) &\longrightarrow& A(t+1)\\
\nonumber   \vec{a}(t) &\longrightarrow& {\vec a}(t+1) = C .{\vec a}(t)\\
\nonumber   f(t) &\longrightarrow& f(t+1) = f(t) - ( {\vec z}(t) \cdot {\vec x}(t)  + {\vec x}(t)\cdot L(\vec{x}(t))
\eneqy
where $L (s_1,s_2,...,S_N)^T = (s_2,s_3,...,S_N,0)^T$, and 
\beq
	C = \left(
\begin{array}{cc}
 \Gamma & I_N\\ -I_N & 0 
\end{array}\right), \,\,\,\,\, 	\Gamma = \left(
\begin{array}{ccccccc}
 0 & 1 & 0 & 0 & ... & 0    & 0 \\
 1 & 0 & 1 & 0 & ... & 0   & 0  \\
 0 & 1 & 0 & 1 & 0   & ... & 0 \\
 \vdots & \vdots &\vdots &\vdots &\vdots &\vdots & \vdots \\
 0 & ... & 0 & 0 & 1 & 0 & 1 \\
 0 & ... & 0 & 0 & 0   & 1 & 0 
\end{array}\right).
\eneq 
It is straightforward then to compute, using $\vec{\epsilon_j} = \epsilon_j (-\vec{1},\vec{1})$ for all $j$,
\beqy
\nonumber &&F^2 \left( \bar T_{\vec{\epsilon}}\right)^{-1}F^2\left[X(u)Z(v)\right]_l F^2\left( \bar T_{\vec{\epsilon}}\right)F^2
  \\
\nonumber &=&F^2\left( \bar T_{\vec{\epsilon}}\right)^{-1}\left[X(-u)Z(-v)\right]_l \prod_{m=0}^{N+1} P(\epsilon_{N+1-m})(m) F^2\\
&=&\label{phase} \zeta^{-\sum_{m=0}^{N+1} ( v[C^m \vec{\epsilon}_{N+1-m}]_{x_l} - u[C^m \vec{\epsilon}_{N+1-m}]_{z_l} )} \left[X(u)Z(v)\right]_l.
\eneqy
This equation shows us that we can analyze the situation for a pulse at a single time $m$. We can ask ourselves what happens when we apply a pulse on time step $N+1-m$: the phase factor in \eqref{phase}
\beqy
&&\nonumber \zeta^{-v[C^m \vec{\epsilon}_{N+1-m}]_{x_l} + u[C^m \vec{\epsilon}_{N+1-m}]_{z_l}}\\
&=&   \zeta^{- v[\vec{\epsilon}_{N+1-m}(m)]_{x_l} - u[\vec{\epsilon}_{N+1-m}(m-1)]_{x_l}}\\
\nonumber &=&\label{solution} \zeta^{ -v \epsilon^{N+1-m} (\theta(l+m-N-1) - \theta(l-m-1))-  u \epsilon^{N+1-m} (\theta(l+m-N-2) - \theta(l-m))}.
\eneqy
To get~\eqref{solution}, we have used that $\vec{a}(t+1) = (\vec{x}(t+1),\vec{z}(t+1)) = (\Gamma \vec{x}(t) + \vec{z}(t), - \vec{x}(t))$  which with the boundary condition $(\vec{x}(0),\vec{z}(0)) = (-\vec{1}, \vec{1})$ yields \beq
[\vec{x}(t)]_l = \theta(l+t-N-1) - \theta(l-t-1).\eneq This means that a pulse in time $N+1-m$ induces the transformation
\beqy
\nonumber X(u)Z(v) &\rightarrow& \zeta^{\epsilon^{N+1-m} u} X(u)Z(v)  \,\,\,\,\,\,  \textrm{ for sites } \{m, N+1-m\},\\
\nonumber &\rightarrow& \zeta^{\epsilon^{N+1-m} (v+u)} X(u)Z(v) \,\,\,\,\,\,\textrm{ for sites in} \,\,\,[m+1, N-m],\\
\nonumber &\rightarrow&  X(u)Z(v)  \,\,\,\,\,\,\textrm   {otherwise.}
\eneqy

Our task is then to find a set of pulses which can generate actions on sites $l$ and $N+1-l$ only. During our calculation we found it helpful to write a sample vector for every time step. Before proceeding we introduce some notation: we will call a $i$-plateau, $M_i(r)$ , a vector with equal non-vanishing components between row $i$ and $N+1-i$ only, additionally we will call an $i$-peak, $R_i(r)$ as a vector with equal non-vanishing elements on sites $i$ and $N+1-m$ only. Thus $M_i(r) - M_{i+1}(r) = R_i(r)$, $M_i(r) = M_i(\alpha r)$ and $M_i(r) + M_i(r') = M_i(r+r')$. 
So not to get the main argument lost in the subsequent section we want to restate what we seek: the problem is solved if we find a set of pulses reproducing an $i$-peak for any $i$ as then Eq.~\eqref{ideal} follows. Thus showing that universal Quantum computation is possible in a QCA fashion. We now proceed to show the solution.

\subsection{The solution:}
Rather than try finding an elaborated numerical formula we shall follow the behavior for sample cases and from there infer the result, in fact the elements we depict are equivalent to take window of the $..., N/2\pm2, N/2\pm1, N/2$ steps of a N element chain.
Lets initially consider the even N case (e.g. $N=8$), with the notation $-v\vec{x} + u \vec{z} = \vec{f}$, 
\beq
\nonumber
\begin{array}{ll} 
\epsilon(t) =&  \left[
\begin{array}{c||c|c|c|c|cc}
S_t & S_0  & S_1 & S_2 & S_{3} & S_{4}  \\
\hline
x_1(t) &-1& 0 & 0 & 0 & 0   \\
x_2(t) &-1& -1 & 0 & 0 & 0   \\
x_3(t) &-1& -1 & -1 & 0 & 0   \\
x_4(t) &-1& -1 & -1 & -1 & 0   \\
x_5(t) &-1& -1 & -1 & -1 & 0   \\
x_6(t) &-1& -1 & -1 & 0 & 0   \\
x_7(t) &-1& -1 & 0 & 0 & 0   \\
x_8(t) &-1& 0 & 0 & 0 & 0   \\
\hline
z_1(t) &1& 1 & 0 & 0 & 0   \\
z_2(t) &1& 1 & 1 & 0 & 0   \\
z_3(t) &1& 1 & 1 & 1 & 0   \\
z_4(t) &1& 1 & 1 & 1 & 1   \\
z_5(t) &1& 1 & 1 & 1 & 1   \\
z_6(t) &1& 1 & 1 & 1 & 0   \\
z_7(t) &1& 1 & 1 & 0 & 0   \\
z_8(t) &1& 1 & 0 & 0 & 0   \\
\end{array}\right] \quad \Rightarrow \\
\\

f =&  \left[
\begin{array}{c||c|c|c|c|cc}
S_t & S_0  & S_1 & S_2& S_{3} & S_{4}  \\  
\hline  
f_1(t) &(u+v)& u & 0 & 0 & 0   \\
f_2(t) &(u+v)& (u+v) & u & 0 & 0   \\
f_3(t) &(u+v)& (u+v) & (u+v) & u & 0   \\
f_4(t) &(u+v)& (u+v) & (u+v) & (u+v) & u   \\
f_5(t) &(u+v)& (u+v) & (u+v) & (u+v) & u   \\
f_6(t) &(u+v)& (u+v) & (u+v) & u & 0   \\
f_7(t) &(u+v)& (u+v) & u & 0 & 0   \\
f_8(t) &(u+v)& u & 0 & 0 & 0   \\
\end{array}
\right]
\end{array}
\eneq
It is important to discriminate at this point between CV and the qudit case: when we have $\epsilon S_i$  then $\epsilon$ can only take values within $\{0,d-1\}$ for the qudit case and $\in \mathbb{R}$ for CV. On CV our calculations are quite simple as we have such freedom that we can choose $\epsilon = 2 \pi/(u+v)$ such that $\zeta^{[\epsilon S_l]}  = \zeta^{\epsilon (\delta_{l} + \delta_{N+1-l})}$ (turning $S_l$ into a $R_l$), however in the qudit case it is not always possible. Nevertheless we can choose to apply pulses in different time-steps with different intensities (value of $\epsilon$) to get the desired result and thus in our example, 
\beqy
\nonumber R_4(u)&=& S_4= u (\delta_4 + \delta_5)\\
\nonumber R_3(u^2)&=& u S_3 - (u+v) S_4 = u^2 (\delta_3 + \delta_6) \\
\nonumber R_2(u^3) &=& u^2 S_2 - (u+v)(u S_3 - v S_4) = u^3 (\delta_2 + \delta_7) \\
\nonumber R_1 (u^{4}) &=& u^3  S_1 - (u+v) ( -u^2 S_2 - vu S_3 + v^2 S_4) = u^4 (\delta_1 + \delta_8)\\
\eneqy
now this wouldn't be successful if $u^s = d \mod d$ for some $s\in \{1,2,...,N/2\}$. However if such is the case then our analysis can be simplified and choosing convenient intensities we would have, say $u^s =d$, 
\beq
\nonumber
R_{i+1}(u^{s-1} v) =  u^{s-1} S_i - u^{s-1} S_{i+1} = u^{s-1} v (\delta_{i+1} + \delta_{N+1-i}) 
\eneq

Now the odd N case, 
\beq
\nonumber
\begin{array} {ll} 
\epsilon(t) =&  \left[
\begin{array}{c||c|c|c|c|cc}
S_t & S_0  & S_1 & S_2& S_{3} & S_{4} \\ 
\hline
x_1(t) &-1& 0 & 0 & 0 & 0   \\
x_2(t) &-1& -1 & 0 & 0 & 0   \\
x_3(t) &-1& -1 & -1 & 0 & 0   \\
x_4(t) &-1& -1 & -1 & -1 & 1   \\
x_5(t) &-1& -1 & -1 & 0 & 0   \\
x_6(t) &-1& -1 & 0 & 0 & 0   \\
x_7(t) &-1& 0 & 0 & 0 & 0   \\
\hline
z_1(t) &1& 1 & 0 & 0 & 0   \\
z_2(t) &1& 1 & 1 & 0 & 0   \\
z_3(t) &1& 1 & 1 & 1 & 0   \\
z_4(t) &1& 1 & 1 & 1 & 1   \\
z_5(t) &1& 1 & 1 & 1 & 0   \\
z_6(t) &1& 1 & 1 & 0 & 0   \\
z_7(t) &1& 1 & 0 & 0 & 0   \\
\end{array}\right] \quad  \Rightarrow \\
\\
f =&  \left[
\begin{array}{c||c|c|c|c|cc}
S_t & S_0  & S_1 & S_2& S_{3} & S_{4} \\  
\hline  
f_1(t) &(u+v)& u & 0 & 0 & 0   \\
f_2(t) &(u+v)& (u+v) & u & 0 & 0   \\
f_3(t) &(u+v)& (u+v) & (u+v) & u & 0   \\
f_4(t) &(u+v)& (u+v) & (u+v) & (u+v) & -v+u   \\
f_5(t) &(u+v)& (u+v) & (u+v) & u & 0   \\
f_6(t) &(u+v)& (u+v) & u & 0 & 0   \\
f_7(t) &(u+v)& u & 0 & 0 & 0   \\
\end{array}
\right]
\end{array}
\eneq
This case follows a similar recipe, 
\begin{equation*}
\begin{array}{rl}
R_4 (v-u)&= S_4\\
R_3 (u(v-u))&= (v-u) S_3  - (u+v) S_4  \\
R_2(u^2(v-u))&= u(v-u) S_2 +   (v+u) (-(v-u) S_3 + v S_4) \\
R_1(u^3(v-u))&= u^2(v-u) S_1 + (v+u) [-u(v-u) S_2 \\
 &+ v(v-u)S_3 - v^2 S_4] .
 \end{array}
\end{equation*}
Now the critical case is when $u^s(v-u) = 0 \mod d$ for some $s$, however if that is the case then 
\beq
\nonumber R_2(u^{s-1}v(v-u))=  u^{s-1}(v-u) S_{i} - u^{s-1}(v-u) S_{i+1}. 
\eneq 
as long as $v-u \neq 0 \mod d$. When $u=v \mod d$ then we have the need of a global pulse $S_0$, which wasn't needed in previous cases to be able to generate a $[\frac{N+1}{2}]$-peak, which is the main point to construct the rest of i-peaks, in our case we would have then that 
\beqy
\nonumber  2 R_4(u) &=&  S_0 + 2(S_3 +S_1-S_2) \\
\nonumber  R_3(u)&=& S_3 -2 R_4(u) \\
\nonumber  R_2(u)&=&S_2 - 2 R_4(u) - 2 R_3(u)\\
\nonumber R_1(u)&=&S_1 -2 (R_4 + R_3 + R_2) 
\eneqy 


This shows that we can always find a set of pulses such that 
\begin{equation}
\begin{array}{lll}
V(\alpha,u,v,l) &=& \prod_j e^{ i \alpha /2 ((1-\zeta^{\kappa}) X(u)Z(v) + (1-\zeta^{-\kappa})(X(-u)Z(-v))^\dagger) \Delta(j,l)}\\
&=& \prod_j  e^{i\alpha/2 (1-\cos(\phi)) (X(u) Z(v) + Z(-v) X(-u)) \Delta(j,l)}\\
&& e^{-i\alpha \sin(\phi) i(X(u) Z(v) - Z(-v) X(-u))\Delta(j,l) }
\end{array}
\end{equation}
with $\cos(\phi)=\Re[\zeta^{\kappa}]$ so that all logical sites can be manipulated at will. Note that while some operations may not be executed on the middle q-site (which holds no information), e.g.$ u-v = 0 \mod d; u+v =0 \mod d$, they can still be performed on the rest of the chain, i.e. the computationally valuable q-sites. Here $\zeta$ and $\kappa$ are not completely freely chosen.  Recall for CV, $\zeta=e^{i}$ and for qudits $\zeta=e^{i2\pi/d}$.  The parameter $\kappa$ is a function of $u$ and $v$ determined by solving the equations above, and is integer for qudits but can be real for CV.  From~\eqref{ideal}, we would need $[N/2]+1-m$ non-vanishing pulses to construct an $m$-peak ($R_m$), which leads to a total number $ 4(N+2) + [\frac{N}{2}] - 2 m$ global operations. Solving the equations system for particular values of $u$ and $v$ simplifies the problem a lot and thus the number of global operations needed, e.g. for $v=0$, $R_i(u) =   S_i - S_{i+1}$. 

\subsection{Summary}
In the previous subsection we have shown that we are able to perform $V(\alpha, u,v,l)$
and thus for whatever the value of $\kappa(u,v)$ we can run the protocol twice to either eliminate or reinforce the Hamiltonian $(X(u)Z(v) + Z(v)^{\dagger}X(u)^{\dagger})$ noting that $1-\zeta^\kappa-\zeta^{-\kappa} \in \mathbb{R}$ and $\Re(\zeta^\kappa-\zeta^{-\kappa}) =0$. This allow us then to apply any element of our operator basis on any logical site, thus achieving single qudit (qunat) arbitrary unitaries. To have complete universal quantum computation we only need now the ability to perform localizable entangling gates. 

To do so we can use a similar methodology to the one used in~\cite{Raussendorf1} for qubits. We can add $\ket{0}$ qudits (qunats) as ancillas between logical sites, preserving the mirror symmetry of the encoding and thus preserving our previous results, such that a T-conjugated $X_i$, $T X^+(s)_i T^{-1} = X(s)_{i-1} \otimes  Z(-s)_{i} \otimes X(s)_{i+1} + X^\dagger(s)_{i-1} \otimes  Z^\dagger(-s)_{i} \otimes X^\dagger(s)_{i+1}$ yields an effective  logical $e^{i(X_{[i]}\otimes X_{[i+1]}+X^\dagger_{[i]}\otimes X^\dagger_{[i+1]})}$ (or entangling gate (as $\ket{0}$ is stabilized by $Z(s)$). 

An alternate and more efficient way of implementing the entangling gate is the following: we perform the sequence to execute an $X^+(u)$ rotation on logical site one, $e^{i \alpha(X(u) + X(u)^\dagger)_{1}}$, then a time-shift
\beqy
\nonumber T^{m} e^{i \alpha(X(u) + X(u)^\dagger)_{1}} T^{-m} \rightarrow  e^{i \alpha(Z(u)_m \otimes X(u)_{m+1} + Z(u)^\dagger_m \otimes X(u)^\dagger_{m+1})}
\eneqy
creates the required entangling gate between nearest neighbors. Note that this method does not need the extra $\ket{0}$ ancillas, therefore reducing the overall number of gates needed to perform two q-site entangling gate. 

To show that this gate is enough for universal quantum computation we can just follow the argument in \cite{LloydBraunstein} for CV. For the qudit case we follow Ref. \cite{Brylinski}, wherein the authors show that if one has a gate set consisting of arbitrary single qudit unitaries and a gate which is diagonal in the computational basis, $V \ket{jk} = e^{i\theta_{jk}} \ket{jk}$, which satisfies  
\beq 
\label{univer}
\theta_{jk} + \theta_{pq} \neq \theta_{jq} + \theta_{pk}, \mod 2\pi\,\,\, \textrm{for some}\,\,\, j,k,p,q,
\eneq 
then this suffices for exactly universal quantum computation.  Exact universality means that any unitary evolution can be obtained by a finite sequence of gates, as opposed to a dense of set of unitaries that cover the group. At our disposal we have $U_{XX}=e^{i(X_{[i]}\otimes X_{[i+1]}+X^\dagger_{[i]}\otimes X^\dagger_{[i+1]})}$ (modulo single q-site unitaries), so we can have it's Fourier conjugated $U_{ZZ}=e^{i(Z_{[i]}\otimes Z_{[i+1]}+Z^\dagger_{[i]}\otimes Z^\dagger_{[i+1]})}$ which has an action $U_{ZZ}\ket{jk} = e^{i 2 \cos(\frac{2\pi}{d}( j+k))}\ket{jk}$. Thus \eqref{univer} turns into $\cos{\frac{2\pi}{d}(j+k)} + \cos{\frac{2\pi}{d}(p+q)} \neq \cos{\frac{2\pi}{d}(j+q)} + \cos{\frac{2\pi}{d}(p+k)} \mod 2\pi$, so for $j = -k = 1$ and $ -p = q = s $ we verify that    
\beqy
\nonumber 2 &\neq& 2 \cos{\frac{2(1+s)\pi}{d}}\,\,\, \textrm{for} \,\,\,s \neq -1, \mod d.
\eneqy
Thus we have universal quantum computation.   

\section{Initialization and Readout}
\label{IO}
Readout of information in the chain can be done in an architecture independent manner using auxiliary states for each subsystem. First we consider the case of qudits ($d$ finite).  We employ two ancillary levels per subsystem, a level $\ket{a}$ that  can be coherently coupled to at least one of the other $d$ information carrying states and another level $\ket{e}$ that couples to the state $\ket{a}$ but none of the information carrying states.  The state $\ket{e}$ should couple to environmental degrees to allow measurement by a classical readout and could represent, e.g. an optically excited state of an atom which decays emitting photons.  High efficiency measurement of population in state $\ket{a}$ is possible if there is a closed cycling transition $\ket{a}\leftrightarrow\ket{e}$.  To realize measurement on any qudit we simply adapt the formalism above but with all the $\tilde{X}_j$ and $\tilde{Z}_j$ operators acting on the $d+1$ dimensional subsystem spanned by $\{\ket{0}_j,\ket{1}_j,\ldots \ket{d-1}_j,\ket{a}_j\}$.  Then to measure population in the logical state $\ket{d-k}_i$ of the qudit located at position $i$ we apply the operator $X^k_i$, so that only the qudit located at site $i$ could have amplitude in an $\ket{a}$ state and then apply a uniform pulse to couple $\ket{a}\rightarrow\ket{e}$ and observe the presence of absence of a classical measurement output.  By composition of these operations readout on any computational state can be performed.
  
For measurement of CVs we could employ an additional degree of freedom per particle.  Say we encode information in the $x$ harmonic oscillator mode, and we have access to control on a $y$ harmonic oscillator mode initially prepared in the vacuum state for readout.  The idea is to swap the state of the $x$ mode (the data) of one CV into its $y$ mode (the ancilla) and perform a global tomographic measurement on moments of the $y$ modes of the CVs.    
This can be done in fact for CVs or for qudits using only global control and global measurements (see Fig.\ref{fig:1}). Our scenario is the following: (i) we use an ancilla chain initialized in the $\ket{0}$ state and (ii) demand that we can let the two chains (computational and ancilla) interact through a $CZ$ gate (site $i$ of the main computational chain with site $i$ of the ancilla chain for all sites $i$). With such conditions we can execute a swap gate between the q-site $l$ of both chains, $Swap_{d_l,a_l} = (F_{d_l} \otimes F_{a_l}) CZ_{d,a} (F^{-1}_{d_l} \otimes F^{-1}_{a_l}) CZ_{d,a} (F^{-1}_{d_l} \otimes F^{-1}_{a_l}) $ as shown in Fig.~\ref{fig:1}.
We have assumed that the ancillary CVs are prepared in the $\ket{0}$ state which are $+1$ eigenstates of the $Z(s)$ operators.  Such states are infinitely squeezed position eigenstates and are not physical (i.e. not normalizable).  We can however achieve highly squeezed states by physically allowed global operations perhaps using the assistance of the associated data CV.  Examples of such protocols include using using extremely short pulses of a standing wave potential \cite{Leibscher} confining the spatial degree of freedom of the CV as demonstrated using trapped atoms in optical dipole potentials \cite{Raizen}

We have considered here another chain, but we have purposely chosen only one of the chains to be  capable of universally controllable such that we can also think of every q-site having two degrees of freedom, e.g. an oscillator in $y$ and $x$, but considering that q-sites can only be coupled though one of the degrees of freedom.

\section{Conclusions}
\label{conc}
In conclusion we have designed a protocol for universal quantum computation with global operations for subsystems encoding arbitrary finite or continuous variables.  The data is stored in a mirror symmetric state over $N$ subsystems encoding $M$ q-sites aligned in one spatial dimension.  The overall requirements in our scheme are: (i) global nearest neighbors $CZ$ gates, (ii) global Fourier pulses, (iii) global $P(\epsilon_m) = X(-\epsilon)Z(\epsilon)$ pulses (iv) the set of Hamiltonians $\{ \Re[X(a)Z(b)]\}$. For quantum computation with CV, condition (i) is met with the homogenous coupling $H=g_1 \sum_{i=1}^{N-1} q_i\otimes q_{i+1}$; condition (ii) with $H=g_2 \sum_{i=1}^{N} (q_i^2+p_i^2)$; condition (iii) is met with $H=\sum_{i=1}^{N}g_3(q_i+p_i)$, and condition (iv) is met with Hamiltonians from the set $\{ \cos{ \omega \hat q}, \sin{ \omega \hat q} \}$ with  $\omega \in \mathbb{R}$.  
We have also proposed a general scheme for readout using only global operations/measurements.

\acknowledgments
We thank DEST ISL Grant CG090188 (JT) and the Mazda Foundation for Arts and Science (GAPS).

\end{document}